# Pressure effects on bond lengths and shape of zigzag single-walled carbon nanotubes

Ali Nasir Imtani and V. K. Jindal[1]

Department of Physics, Panjab University, Chandigarh-160014, India

We investigate structural parameters, i.e., bond lengths and bond angles of isolated uncapped zigzag single-wall nanotubes in detail. The bond lengths and bond angles are determined for several radii tubes by using a theoretical procedure based on the helical and rotational symmetry for atom coordinates generation, coupled with Tersoff potential for interaction energy calculations. Results show that the structure of zigzag tubes is governed by two bond lengths. One bond length is found to have a value equal to that of graphite, while the other one is larger. Furthermore, the tube length is found to have significant effect only on larger bond length in zigzag tubes. With the application of the pressure, only the larger bond length compresses, the other one remaining practically constant. At some critical pressure, this bond length becomes equal to constant bond length. This behavior of bond lengths is different from those of armchair tubes. An analysis regarding the cross sectional shape has also been done. At some higher pressure, transition from circular to oval cross section takes place. This transition pressure is found to be equal 2.06GPa for (20,0) tube. Some comparison with chiral tubes has also been made and important differences on bond length behavior have been observed.

## I. Introduction

Carbon nanotubes (CNTs) are fascinating materials whose mechanical and transport properties are studied at length at nanoscale. The unique structure of single-wall nanotubes (SWNTs) in the form of a cylinder by rolling a graphite sheet with a particular wrapping angle is responsible for observed electronic properties, such as conductivity [1]. The stiff and high-enthalpy C-C bonds in the structure also determine excellent mechanical properties such as high tensile strength [2,3]. Changes in atomic configuration can manifest themselves as local structure deformations in the bonds. Such structural changes can be induced by various chemical treatments or by mechanical means such as exerting stress, bending strain or through application of pressure on the tube, which can be seen as probes. A matter of significant interest has been how these properties alter when the wrapping angle or twist is changed.

In this work we have concentrated on looking at the behavior of the bond lengths and bond angles for uncapped isolated zigzag SWNTs as a function of zigzag tube dimensions. Ideally, a nanotube is constructed by rolling up a perfect graphene sheet [4]. The C-C bond length is the only parameter to determine the structure. For graphite sheet this bond length is 1.42 Å [5]. It is common practice to choose this value of bond length as that for nanotubes also. However, in our earlier work [6] while studying armchair SWNTs we observed that two bond lengths need to be considered for describing the structure of SWNTs. There are many other theoretical studies devoted to the variation of the bond lengths in zigzag SWNTs [7-10]. They have also reported the results of two bond lengths. Kurti et al. [7], who used local density approximation (LDA) to determine the radial breathing mode (RBM) frequency of various zigzag and armchair tubes with radii between 3.5 Å and 8.1 Å, found that the carbon-carbon bond lengths are not uniform in the tubes. Gulseren et al. [8], who investigated the curvature effects on geometric parameters, energetics, and electronic structure of zigzag single-walled nanotubes with fully optimized geometries from first-principle calculations, found that both bond lengths and the bond angles display a monotonic variation and approach the graphene values as the radius increases. They also observed that one bond length is smaller than the graphite value while the other one is larger. Other results [9] show that the nanotube radius has a little effect on the mechanical behavior of single-walled carbon nanotubes subject to simple tension or pure torsion, while the nanotubes orientation has somewhat larger influences. They also reported numerical results of the bond lengths, bond angles and carbon nanotube diameter. Ab initio molecular dynamics simulations have also been performed to investigate the equilibrium carbon-carbon bond lengths and bond angles of small radii single walled nanotubes [10]. They showed that for both zigzag and armchair nanotubes there are two nonequivalent bond lengths and small variation on the bond angles. The short range Tersoff's potential with the extensive molecular-dynamics simulations has been successfully applied to study the structural characteristics of the lattice SWNTs[11].

---

[1] Author with whom correspondence be made, e-mail: jindal@pu.ac.in

A study of the structural and mechanical properties of carbon nanotubes under pressure provides further interesting results. A number of experiments have been carried out on bundles of SWNTs [12-14, 16,18-19] at high pressure, showing pressure induced structural transitions in the range of 1-2 GPa [12-14]. Experiments showed that the pressure may induce transitions in electrical and magnetotransport properties in SWNT bundles[17] and individualized SWTNTs[15], which correlated closely with the pressure induced structural shape transition [17]. Recent experimental study of the high pressure behavior of a bundle of $1.35 \pm 0.1$ nm diameter single wall carbon nanotubes filled with fullerenes has been performed by Raman spectroscopy [20]. They show that two reversible pressure induced transitions take place in the compressed bundle SWNTs. The first transition occurs at about 2-2.5 GPa independent of the choice of the pressure transmitting medium, as well as of the filling or not of the nanotubes. These results are close to a recent calculation on armchair carbon nanotubes[6].

Extensive theoretical studies under hydrostatic pressure using first-principle calculation and molecular dynamics simulations [21-28] have also been performed by several groups to study the properties of both isolated and bundles of SWNTs. So far theoretical studies on isolated single wall nanotubes under pressure have mostly focused on armchair and zigzag tubes [6,23-25], which have a high symmetrical radial atomic structure and a short axial period. It has been found that the pressure induces a series of shape transitions in both armchair and zigzag SWNTs. It must be noticed that a study of graphite under pressure has shown that for pressures higher that 17 GPa graphite undergoes a phase transition with bonding changes and bridging carbon atoms between graphite layers [29].

As a result of all this theoretical analysis, it seems that no coherent picture has emerged about the behavior of bond lengths in zigzag tubes, calling for a comprehensive study on the behavior of bond lengths in these tubes of varying tube radii and length. In order to have an insight into pressure induced structural transformation, it is preferable to have a detailed study based on suitable model potential. So it is necessary to understand the behavior of bond lengths under pressure for which no detailed study exists. This paper extends the results of our earlier work on armchair tubes [6] to zigzag (n,0) SWNTs, where n=5, 10, 15, 20, 30, 40, 50 and 70. The behavior of bond lengths has been studied in detailed way. A typical isolated zigzag SWNT is shown in Fig.1. The bond length $b_1$ is parallel to the tube axis whereas $b_2$ forms an angle with it. The length $L$ of the tube is determined by fixing the number of unit cells, $N$, which make the length of all tubes approximately equal. We can calculate the length of zigzag SWNTs in terms of bond lengths $b_1$ and $b_2$ by $L = (N(b_1 + 2b_2) - b_2)$. The number of atoms in each tube then equals $4Nn$ for (n,0) tubes, different for different radii. We have taken $N$ equal to 29 for our calculation in this work.

In the next section, we describe our method to obtain structure of zigzag tubes based on two different bond lengths. This has been used to calculate minimum energy structure using Tersoff potential. Results thus calculated are presented in Section III. Effect of hydrostatic pressure has been discussed in Section IV and summary and conclusion has been given in Section V.

II. **Helical and rotational symmetries**

We can visualize an infinite tube as a conformal mapping of a two-dimensional honeycomb lattice to the surface of a cylinder that is subject to periodic boundaries both around the cylinder and along its axis. First, we assume that the cross section of isolated tubes has a circular shape. The helical and rotational symmetries have been used earlier to generate atomic coordinates for nanotubes used equal bond lengths [30]. We modify this procedure to obtain atomic coordinates using two different bond lengths for high symmetry zigzag (n,0) SWNT's (with $\theta = 0^o$)(Fig.1). This is done by first mapping the two atoms in the [0,0] unit cell to the surface of cylindrical shape. The first atom is mapped to an arbitrary point on the cylindrical surface [e.g., $(R,0,0)$], where $R$ is the tube radius in terms of bond length $b_2$ and the position of the second atom is found by rotating this point by angle $\phi = \pi/n$ about the cylinder axis in conjunction with translating it by the distance $h_t = \frac{1}{2}b_2$. These first two atoms can be used to locate $2(n-1)$ additional atoms on the cylindrical surface by $(n-1)$ successive $2\pi/n$ rotations about the cylinder axis. Altogether, these 2n atoms complete the specification of the helical motif which corresponding to an area on the cylindrical surface. This helical can then be used to tile the reminder of the tubule by repeated operation of the single screw operation $S(h,\alpha_h)$ representing a translation $h$ along the cylinder axis and rotation $\alpha_h$ about this axis, where $h = b_1$ and $\alpha_h = \pi/n$.

If we apply the full helical motif, then the entire structure of zigzag SWNT is generated. This structure provides the atomic position of all the atoms in terms of bond lengths. The bond lengths are determined by minimization of the energy of the tube, assuming atoms interact via Tersoff potential [31].

### III- Results and Discussion
### A-Equilibrium shape of the cross section

We study two sets of the initial structures, (1) perfect circular cross section and (2) elliptical cross section with different aspect ratio $b_e/a_e$, where $a_e$ and $b_e$ are the longer and shorter radius, respectively. The lengths of the tubes are taken from $L = (N(b_1 + 2b_2) - b_2)$ for N=29 come out to be to be slightly longer than 120Å. The reason to take this length is that the C-C bond lengths and bond angles of zigzag tubes attain constant values at this length and thus this length can be described as long tube length (see section C). We adopted the same procedure, as described in [6], to construct the single-walled nanotubes with circular and elliptical cross sections and then minimizing the energy using Tersoff potential. Fig. 2 shows the variation of the energy thus calculated for zigzag (n,0) SWNTs. We observe that the energy for all the tubes with circular cross section is lower than the energy with elliptical cross section. The deviation increases with decrease in $b_e/a_e$. This result indicates that the equilibrium shape of the cross section of zigzag tubes is a circular shape.

**B- Effect of tube radius on the structure**

In Table I, we present, the results of our calculations for two different bond lengths and three bond angles for several zigzag SWNTs. The normalized bond lengths (i.e. $b_{1,2}/b_o$ where $b_o$ is the bond length in graphite sheet) and bond angles as obtained by us have been plotted as a function of the tube radius in Fig.3(a) and Fig.3(b) respectively. The difference between one bond length $b_2$ and bond angles in SWNTs from those in graphite is significant. Form Fig.3(a), we observe that $b_2$ is larger than of graphitic value for all zigzag tubes, while $b_1$ has a value equal to that of graphite. For very small radii tube such as (5,0) tube, $b_1$ becomes smaller than of graphitic value. In case of armchair tubes, also we found that two bond lengths determine the structure [6]. In that case, one of them was larger than that of the graphitic value while the other was smaller than it.

We now compare our results for bond lengths for several radii zigzag tubes (presented in Table I) with other calculations [7-10]. These calculations also found two bond lengths in the structure of zigzag tubes, but the bond length $b_1$ was found to be smaller than the graphite value whereas $b_2$ was found to be larger [8,10]. While in other studies these two bond lengths have values larger than that of graphite value [9], but in [7] these are smaller. They obtained the bond lengths $b_1$ and $b_2$ for (5,0) tubes as 1.4669 Å and 1.4542 Å and for (10,0) tubes as 1.4544 Å and 1.4518 Å[9], respectively. For large radii tubes such as (30,0) tube, the bond lengths are 1.4511 Å and 1.4508 Å [9]. In [7], for all zigzag tubes with radii between 3.5 Å and 8.1 Å these bond lengths are equal to $b_1$=1.408 Å and $b_2$=1.413 Å. Our results of the tubes radii are relatively closer in agreement with the results obtained by [7] for (10,0), (15,0) and (20,0) tubes. The obtained values of the radii for these tubes are equal to 3.5 Å, 7.8 Å and 5.8 Å, respectively. It should be noticed that some of the calculations [9,10] did not reach a satisfactory graphite bond length from the beginning, these reproducing C-C bond lengths as 1.4507 Å and 1.415 Å, respectively in contrast to 1.42 Å for graphite. We are able to reproduce C-C bond length of 1.42 Å for graphite using the potential parameters as defined in [6].

In general, it is observed that with increase in the tube radius $b_2$ decreases to approach to the graphite value for larger radii tubes. These effects can be correlated with the curvature. The curvature energy $E_c$ as a function of tube radius is shown in Fig. 3(c). It is however interesting to note that even (70,0) tube with radius 27.420Å has about 5.5% curvature energy still remaining.

There are three bond angles in the structure of zigzag tubes. Two of them $\alpha$ and $\gamma$ are equal and larger than the third one $\beta$ (see Table I). Because of non-co-planar nature, their sum is smaller than 360°. This is particularly true for small radii tubes. As the tube radius increases, the bond angles become closer to ideal value ($\approx 120°$). In the case of armchair tubes [6], two of the three bond angles are also equal but smaller than the third one.

We also report results for chiral SWNTs adopting similar procedure to obtain the structural parameters. Preliminary results on these indicate that these structures are characterized by only one bond length. We present in Table II, the results obtained for bond length for chiral tubes having different radii with the same chirality (i.e., the same chiral angle $\theta = 4.175^o$). We also plot the normalized bond length with tube radius in Fig.3 (d). In general, we observe that the tubes with different radii have values of the bond length larger than that of graphite. This bond length changes slightly with the tube radius. On the basis of the comparison of bond lengths of armchair, zigzag and chiral tubes, it seems that curvature effects arise mainly from chirality, radius having a much smaller effect. As an example, the curvature energies of zigzag (50,0) and chiral (50,5) tubes which have approximately the same radius, are 0.0573 and 0.3175 eV/atom, respectively. Detailed results on chiral tubes involving several chiralities will be published separately.

**C- Effect of tube length on the structure**

Length can have significant effect on the structure and electronic properties of SWNTs. Rochefort et al. [32] have studied theoretically the influence of the finite length on the electronic properties of (6,6) SWNT. They found that, unlike the infinite tube, which is metallic, very short (<100Å) nanotubes have an energy band gap. Therefore, we also study of the length effect on the bond lengths of zigzag SWNTs.

We choose three zigzag (10,0), (20,0) and (30,0) single-wall nanotubes for our study. Fig. 4(a) shows the variation of the bond lengths calculated by us (in units of graphite value) plotted as a function of aspect ratio i.e. the length to radius ratio $(L/R)$. We observe that as the tube length increases, the larger bond length $b_2$ increases. At some aspect ratio, it becomes constant. The value of this aspect ratio is $\approx 7.0$ for all tubes. We also observe that $b_2$ is always larger than that of graphite value. For small radii tubes, the difference from the graphite value is larger. The bond length $b_1$ and bond angles remain constant with increase in the tube length. These results are plotted in Fig.4 (a) and Fig.4 (b) for (10,0) tube. The results from other radii zigzag tubes are similar. The length also has important effect on the energy of nanotubes. Increase of the tube length tends to decrease the curvature energy (Fig.4(c)).

We also calculate for comparison the effect of length for chiral tubes, in which case the length has been found to have no effect on the value of the bond length (Fig.4(d)). In the case of armchair tubes, the tube length has much more significant effect on the bond lengths and bond angles [6].

## IV. Pressure effect

In order to calculate the zigzag tube cross-sectional structure under hydrostatic pressure, we first assume the cross-section to be circular and obtain minimum energy and bond lengths at various hydrostatic pressures. Subsequently, we allow change of cross-section to elliptical shape and recalculate structure.

### 1. Circular cross section

The total potential energy $E_o$ under a hydrostatic pressure $P$ of the tube is given by:

$$E = E_0 + P\Delta V, \qquad (1)$$

where $\Delta V = V_p - V_0$, $V_p$ is the volume under applied pressure, $V_0$ and $E_0$ are the volume and energy at zero pressure. For finite atoms on a cylinder, the values of pressure and volume used here represent averaged values using localized forces acting on various atoms. We search about the suitable values of the bond lengths $b_1$ and $b_2$ leading to the minimum energy under pressure $P$, as obtained from Eq.(1). For this, we first fix a new $b_2$ (corresponding to some equivalent pressure P obtainable from $\Delta E / \Delta V$), and obtain $b_1$ from minimized energy $E$. By successive minimization procedure, we calculate a set values of $b_1$ and $b_2$ at any pressure. The bond angles are also obtained at each pressure from the final minimized set of atomic positions. The volume, $V_p$, at each pressure is obtained by assuming those tubes to have a circular or elliptical cross section, where the radius and length are governed by $b_1$ and $b_2$. The volume considered here is based on atomic positions defining a continuous surface – new positions at different pressures defining different volume. Since the calculations are based on minimizing energy, which may be obtained through force-distance or equivalently through pressure-volume analysis. It seems that this definition of volume is actually equivalent to modified distances.

Fig.5 shows the results of our calculations of the bond lengths and bond angles at various pressures for (10,0), (20,0), (30,0) and (50,0) SWNT's. We observe that only the larger bond length $b_2$ compress under pressure. Such kind of the behavior of bond length has also been observed in the case of anisotropic lamellar system in InSe [34]. At some critical values of pressure ($P = P_c$), the larger bond length $b_2$ becomes equal to constant bond length $b_1$. This critical pressure is found to depend on the tube radius (Fig. 5(e)). Above this critical pressure, the larger bond length $b_2$ continues to compress. As expected, it emerges that the value of $P_c$ reduces with increasing tube radius. The critical pressure decreases with increasing radius rather quickly and beyond 19 Å or (50,0) tube, it becomes very small, less than 1 bar. In fact for (50,0) tubes or above the curvature effects reduce significantly as observed earlier that in Fig.3(c). We also observe that the behavior of the bond angles. They behave similar to bond length $b_1$, remaining unaltered under pressure for all zigzag tubes. This is shown in Fig.5(f) (plotted only for (10,0) tube). Since $R = \dfrac{n\sqrt{3}b_2}{2\pi}$, indicating that it depends only on $b_2$, whereas the tube length depends on $b_1$ and $b_2$, this results in an unequal compression in the circumferential and axial directions. This behavior has been shown in Fig.6(a) for (10,0) tube.

This result is in good agreement with several observations made earlier that the single-wall nanotubes are extremely rigid along the tube axis than in the circumferential direction. A better estimate of stiffness of carbon nanotubes is provided by the results of bulk modulus obtained from $B = -V(\Delta P/\Delta V)$. The slope $\Delta P/\Delta V$ is obtained from the $P-V$ curve fitted to a quadratic polynomial. These results of the bulk modulus of zigzag tubes are plotted in Fig.6(b) at zero pressure. There is a strong radius dependence of bulk modulus.

## 2. Shape Transition

For study of change of cross section, we choose three isolated zigzag (20,0), (30,0) and (50,0) SWNTs. Under pressure, the cross section assumes to transform from circular to elliptical shape with $b_e/a_e$ as parameter.

Fig.7(a) shows the energy of zigzag SWNTs as a function of pressure. It contains two parts: AB curve corresponds to a circular cross section while CD curves to elliptical cross section (the collapsed tube). Each point on the CD curve represents a different value of $b_e/a_e$ ($b_e/a_e$ begins from the value 0.98 to 0.999). We observe that at some region of pressure, depending on the tube radius, the energy of tube with elliptical cross section is lower than the energy with circular cross section. This result indicates that the shape transition, from circular to oval cross section, occurs in zigzag tubes. The transition pressure ($P_T$) comes out to be 2.06GPa with $b_e/a_e$ equal to 0.999 for (20, 0) tube. In Fig. 7(b), we have plotted the transition pressure as a function of tube radius for zigzag tubes. By observing this figure, as the tube radius increases the value of the transition pressure decreases. The dependence of the transition pressure on $b_e/a_e$ in addition to tube radius is shown in Fig. 7( c). As $b_e/a_e$ decreases the transition pressure increases. It is interesting to compare the values of the transition pressure of zigzag tube with armchair tube having approximately the same radius. The transition pressure for armchair (10,10) tube with $b_e/a_e$ equal to 0.99 is slightly higher (2.26 GPa[6]) as compare to that for zigzag (20,0) tubes (2.06 GPa) with $b_e/a_e$ = 0.999. In fact, detailed pressure dependent measurement of phonons in carbon nanotubes through Raman spectroscopy manifest this change of the cross section through disappearing of RBM [18, 19].

It should be noticed that the value of the critical pressure ($P_c$) defined earlier representing as critical pressure at which the bond lengths become equal in circular cross section (Fig. 5(e)) is smaller than the value of the transition pressure($P_T$). This presents interesting result, leading us to conclude that at some critical pressure; firstly, the bond lengths become equal in circular cross section and then at some increased transition pressure, the cross section transform to elliptical shape.

Bulk modulus, at the transition pressure, for the collapsed zigzag (20,0) (30,0) and (50,0) SWNTs are found to be equal to 60.01, 42.3 and 27.4 GPa, respectively, reducing drastically from their values in circular cross section. This can be easily seen from Fig. 7(e) where we plot the values of bulk modulus at zero pressure, at pressure equal to $P_T$ but with circular cross section and at $P_T$.

**V- Summary and Conclusion**

In this paper, we have investigated in detail the effect of radius and tube length on the values of C-C bond lengths and bond angles of isolated zigzag single-wall nanotubes. We also present some results for chiral tubes of given chirality. The variation of these C-C bond lengths and bond angles as well as cross sectional shape change under hydrostatic pressure has also been presented in this work.

It emerges that the structure of zigzag tubes has unequal bond lengths and bond angles. One of these bond lengths has a constant value equal to that of graphitic value while the other bond length is larger than this for all zigzag tubes. Two of three bond angles are found to be equal and larger than the third one. The larger bond length depends on the tube length strongly while other one and bond angles remain not change with increase in the tube length. On the other hand, for chiral tubes, only one bond length completely describes their equilibrium structure. Furthermore, in contrast to zigzag armchair tubes, tube length has no effect on this bond length.

We also calculate the structure of zigzag SWNTs under hydrostatic pressure. Only the larger bond length responds to the pressure. It decreases with pressure. At some critical pressure, the larger bond length becomes equal to constant bond length.

We also predict the existence of shape transition, from circular to elliptical cross section at some higher pressure, called transition pressure. It is found that the transition pressure depends on the tube radius in addition to elliptical aspect ratio.

The observations made here regarding differences in bond length behavior for different chirality tubes, of armchair, zigzag and chiral variety, especially under hydrostatic pressure should be exploited for characterizing the type of the tube by experimentally correlating possibly observable critical pressure in (n,m) tubes in Raman data or any other experiment which reflects structural parameters. The phonon modes in Raman experiments could broaden depending upon difference in the bond lengths. There could also be a shift in mode frequencies. A careful observation near the critical as well as transition pressure may be interesting. However, it should be noted that the pressure transmitting medium may mask some observations in the experimental data [33].

We would like to emphasize that the study of individual or individualized nanotubes under high pressure is an important developing route. In particular, the exploration of small diameter individual nanotubes under pressure would permit verification of the results presented in the manuscript. We hope that the results presented here provide enough motivation for possible experimental study to verify the differences of the pressure effects of bond-length distribution in zig-zig and armchair tubes.

Table I: Bond lengths, bond angles, radius (Å) and energy $E$ (eV/atom) for zigzag (n,0) SWNTs.

| SWNT | bond lengths(Å) | | bond angles (deg.) | | | | |
|---|---|---|---|---|---|---|---|
| (n,m) | $b_1$ | $b_2$ | $\alpha$ | $\beta$ | $\gamma$ | Radius | Energy |
| (5,0) | 1.4188 | 1.470 | 120.362 | 110.166 | 120.362 | 2.0261 | -6.8243 |
| (10,0) | 1.4201 | 1.431 | 120.053 | 117.357 | 120.053 | 3.9447 | -7.2202 |
| (15,0) | 1.4203 | 1.425 | 119.997 | 118.784 | 119.997 | 5.8923 | -7.2800 |
| (20,0) | 1.4204 | 1.423 | 119.977 | 119.292 | 119.977 | 7.8454 | -7.2991 |
| (30,0) | 1.42045 | 1.4215 | 119.963 | 119.657 | 119.963 | 11.755 | -7.3120 |
| (40,0) | 1.4205 | 1.4210 | 119.958 | 119.786 | 119.958 | 15.668 | -7.3163 |
| (50,0) | 1.4205 | 1.4210 | 119.955 | 119.845 | 119.955 | 19.585 | -7.3183 |
| (70,0) | 1.4205 | 1.4210 | 119.953 | 119.897 | 119.953 | 27.420 | -7.3201 |

Table II: Radius, bond length and energy $E$ (eV/atom) for chiral (n,m) SWNTs at the same chirality.

| Chiral angle $\theta$=4.1750° | | | |
|---|---|---|---|
| SWNT | Radius(Å) | $b$ (Å) | Energy |
| (10,1) | 4.17348 | 1.4370 | -7.05070 |
| (20,2) | 8.33593 | 1.4351 | -7.05115 |
| (30,3) | 12.5030 | 1.4350 | -7.05415 |
| (40,4) | 16.6649 | 1.4345 | -7.05733 |
| (50,5) | 20.8311 | 1.4345 | -7.05815 |
| (100,10) | 83.3533 | 1.4345 | -7.05911 |

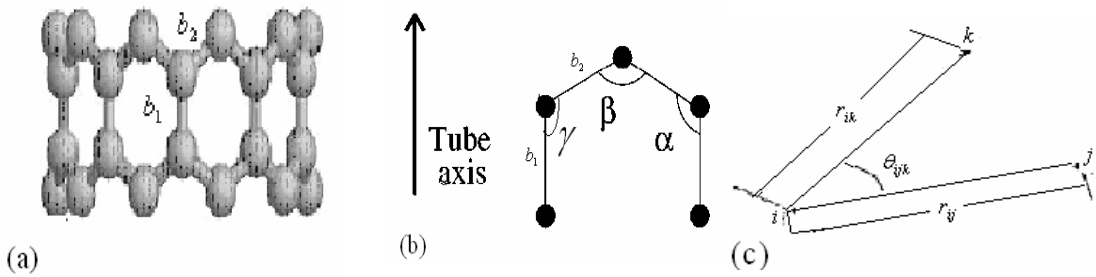

Fig.1: (a) A schematic side view of zigzag SWNT and (b) a part of zigzag SWNT indicating two types of C-C bonds, these are labeled as $b_1$ and $b_2$, and three bond angles $\alpha$, $\beta$ and $\gamma$. (b) Carbon atoms $i$, $j$ and $k$, the corresponding bonds $i-j$ and $i-k$ and bond angles $\theta_{ijk}$.

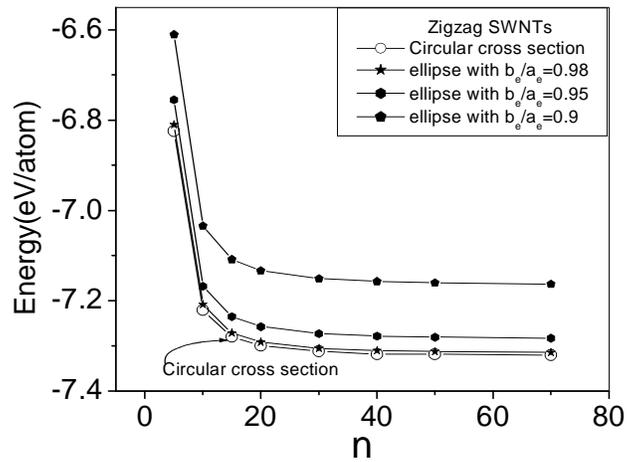

Fig.2: Energy as a function of $n$ for zigzag (n,0) SWNTs having circular cross section ($b_e/a_e$=1.0) and various elliptical cross section (corresponding to different values of $b_e/a_e$), where $b_e$ and $a_e$ are the shorter and longer radii in elliptical cross section.

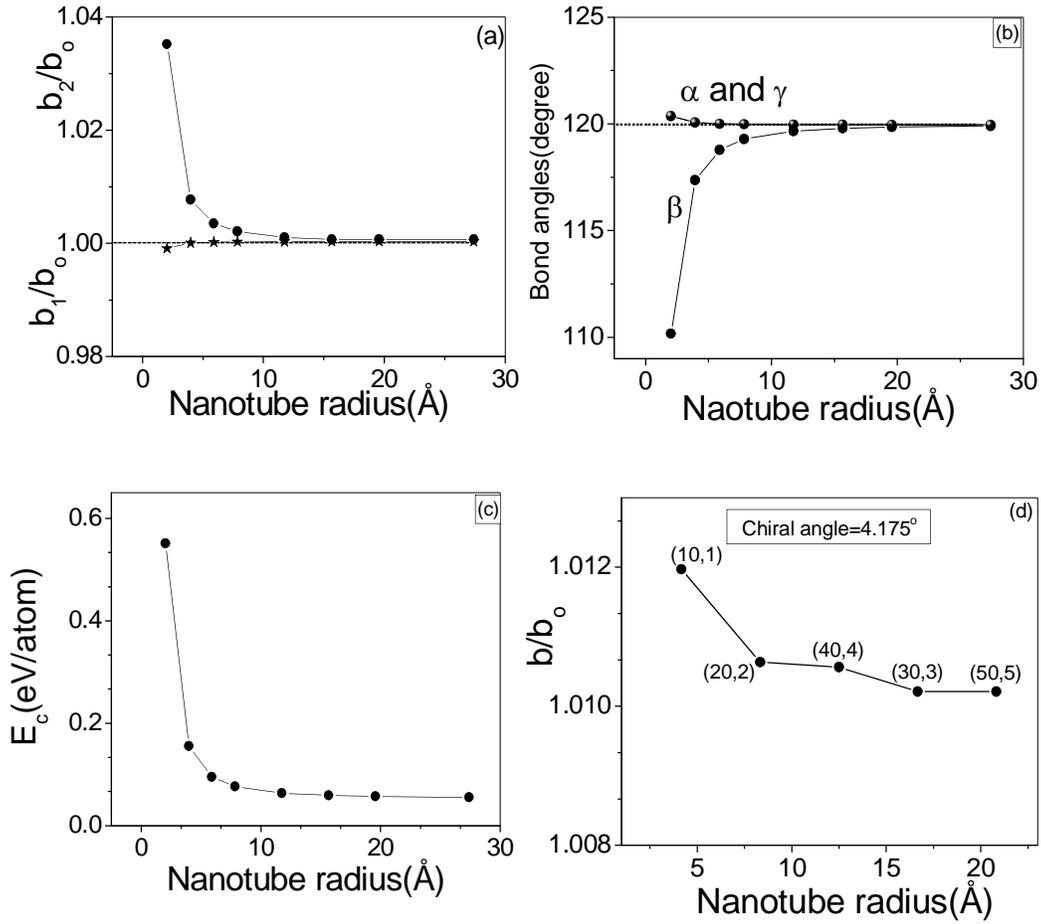

Fig. 3: (a) Normalized bond lengths $b_1/b_o$ and $b_2/b_o$, (b) bond angles α and β, and (c) curvature energy $E_c$ as a function of tube radius for zigzag SWNTs. (d) Variation of normalized bond length with tube radius for chiral tubes. In fig. 3(b), the bond angle γ is equal to α.

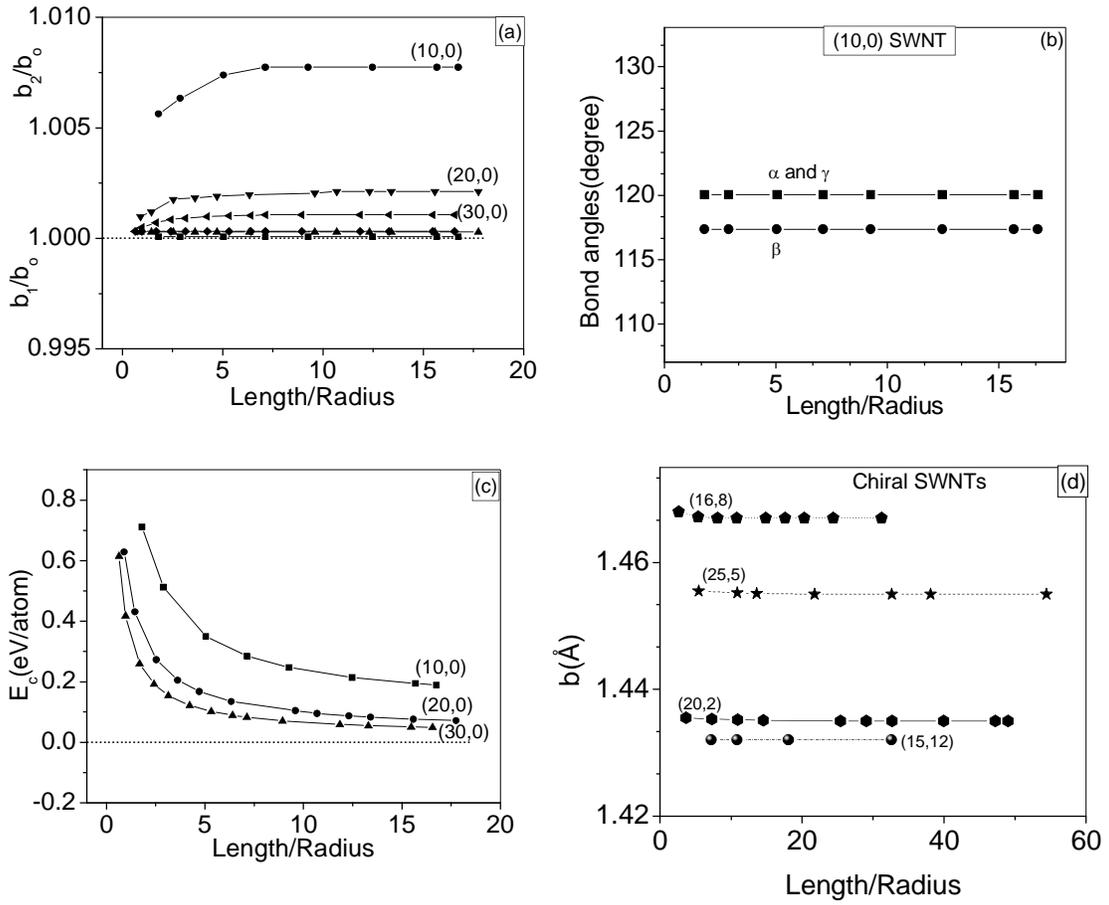

Fig. 4: (a) Normalized bond lengths $b_1/b_0$ and $b_2/b_0$, (b) bond angles for (10,0) tube and (c) curvature energy $E_c$ for zigzag tubes and as a function of the length to radius ratio. Only one bond length $b_2$ affects with tube length. (d) Bond length with length to radius ratio for chiral tubes.

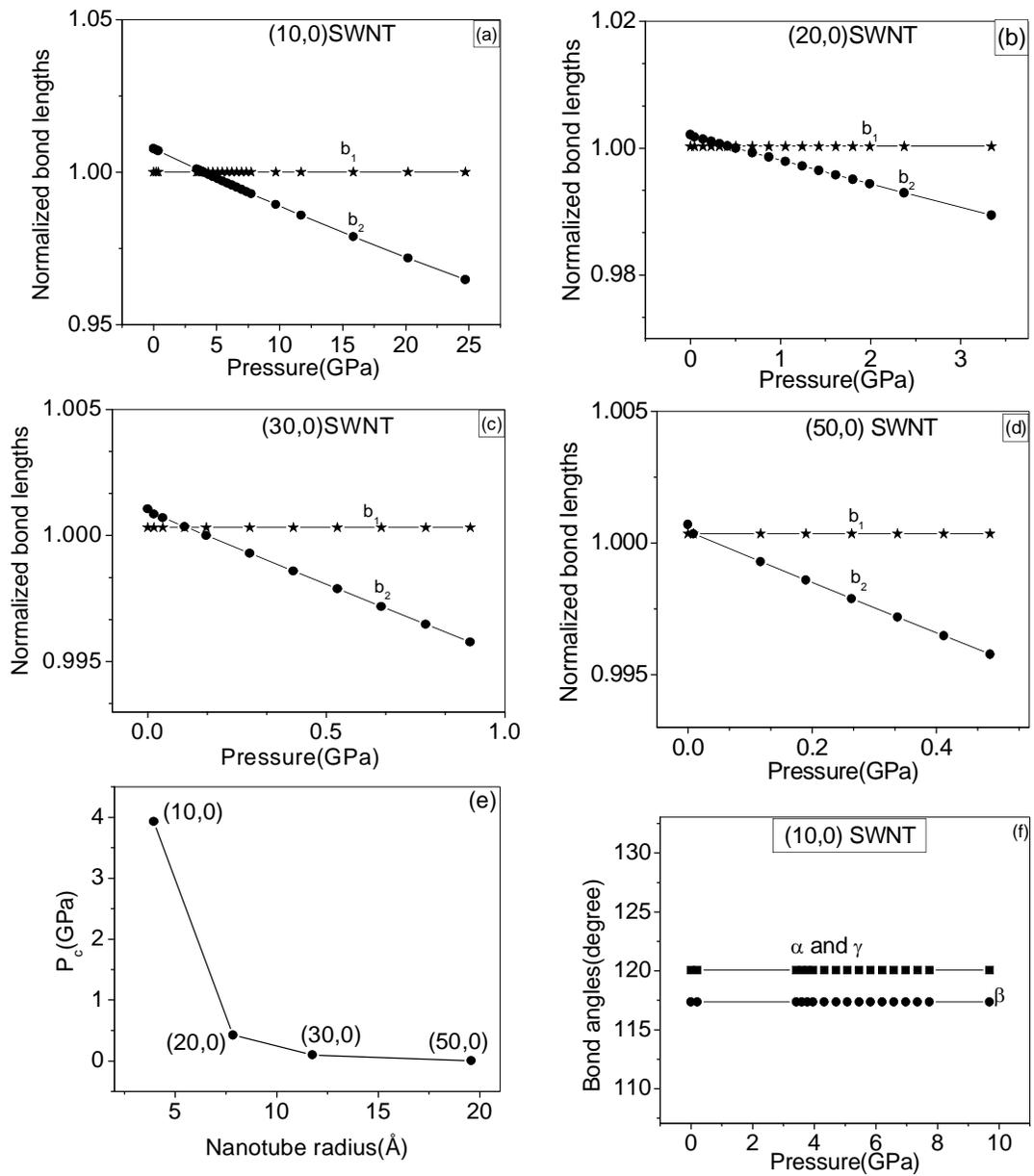

Fig.5: (a) to (d) Normalized bond lengths $b_{1,2}/b_o$ as a function of applied pressure for zigzag (10,0), (20,0), (30,0) and (50,0) SWNTs respectively. (e) Critical pressure $P_c$ versus tube radius and (f) bond angles with pressure for (10,0) tube.

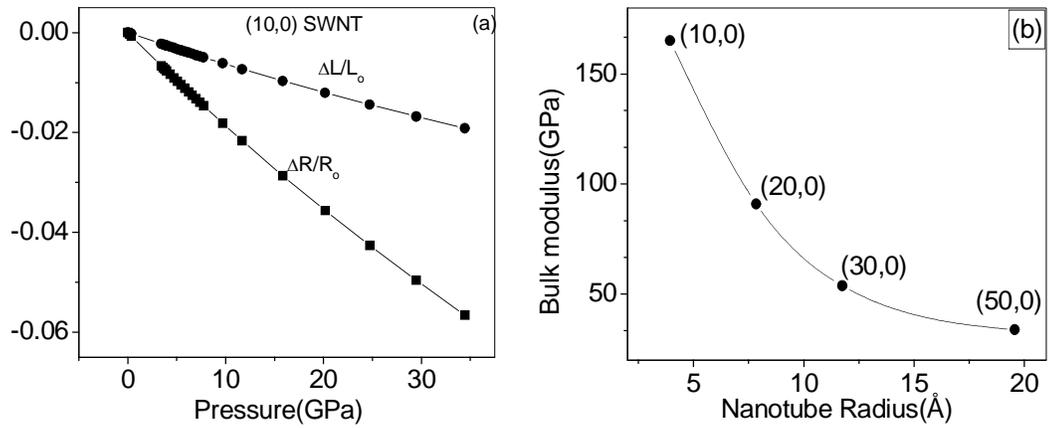

Fig.6: (a) Compression along the tube axis and in the circumferential direction under applied pressure for (10,0) SWNT. (b) Bulk moduls as a function of tube radius at zero pressure for zigzag tubes.

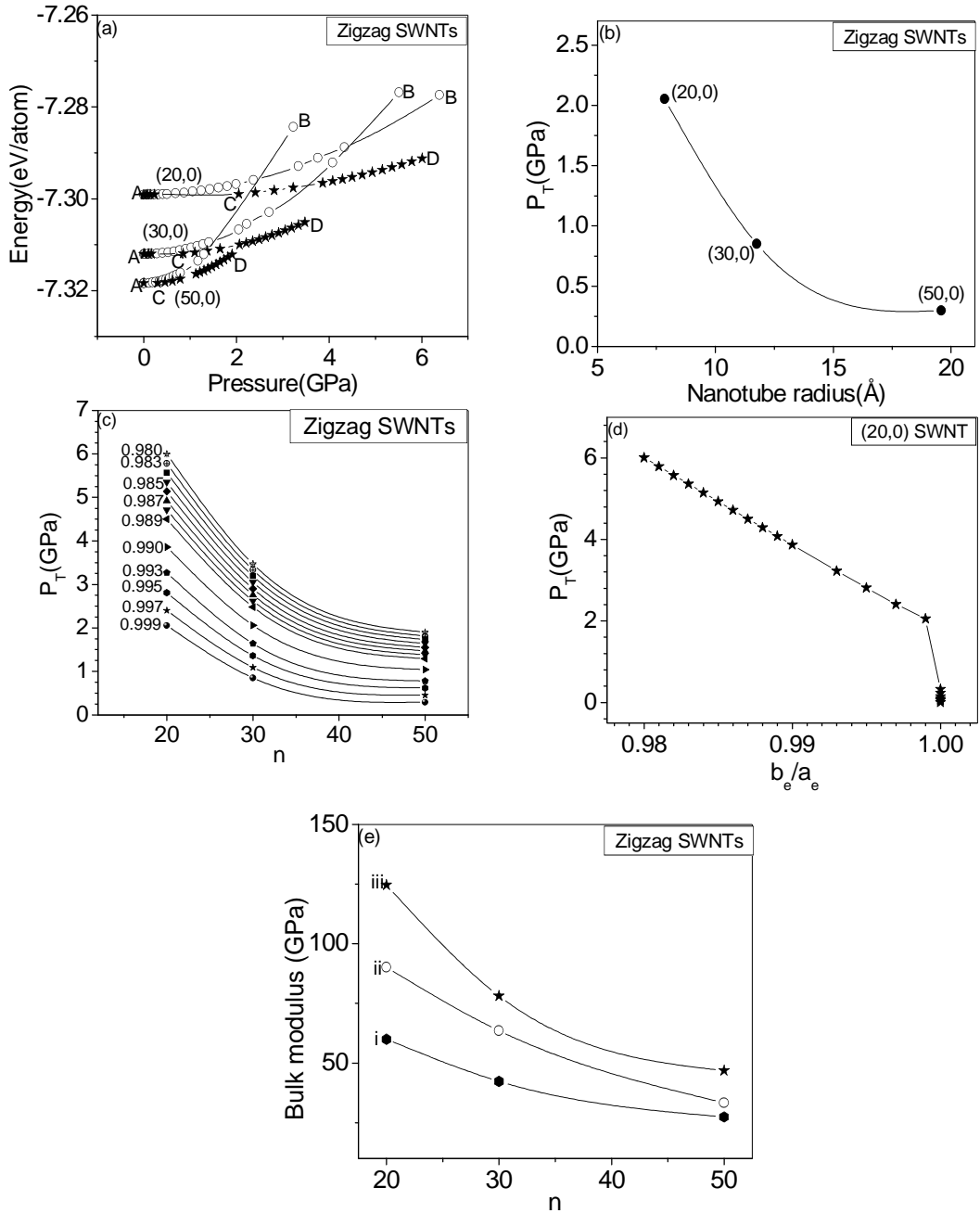

Fig.7: (a) Energy as a function of pressure. AB(open circles) and CD( stars) curves corresponding to circular and elliptical cross section, respectively. (b) Transition pressure versus tube radius at $b_e/a_e$ equal to 0.999, (c) transition pressure with $n$ at different elliptical aspect ratio for zigzag tubes, (d) transition pressure with $b_e/a_e$ for (20,0) tube and (e) bulk modulus versus $n$ for zigzag (n,0) tubes. Curves: (i) at $P_T$, (ii) at zero pressure and (iii) at pressure equal to $P_T$ but with circular cross section.

2121